\documentclass[a4paper,11pt,twocolumn]{article}
\pdfoutput=1 % enforce pdflatex rather than latex for arxiv processing

\usepackage{graphicx}
\usepackage{xcolor}
\usepackage{url}
\usepackage[colorlinks=true,linkcolor=blue,citecolor=blue,urlcolor=blue]{hyperref}
\usepackage[expproduct=cdot]{siunitx} %expproduct=tighttimes
\usepackage{amsmath}
\usepackage{bbold} % for \mathbb{c} (small case)
\usepackage[charter]{mathdesign} % SECOND BEST
\usepackage{ctable} % \toprule, \thickmidrule, \midrule, \cmidrule and \bottomrule to replace \hline in tables
\usepackage{tabularx} % For extensible columns. Alternative: tabulary. Calls \usepackage{array}
\usepackage{cite} % to display references [1, 2, 3] as [1-3] (without using natbib)

\definecolor{royalblue4}{HTML}{27408B}% royalblue4 in Emacs (see M-x list-colors-display)
\definecolor{red4}{HTML}{8B0000}% red4 in Emacs
\definecolor{green4}{HTML}{008b00} % green4 in Emacs

\usepackage{dblfloatfix} % to use modifiers [h] etc. in \begin{figure*} ... \end{figure*} environment
\usepackage{fixltx2e} % prevent starred-figures from appearing out of ascending order

\usepackage[font=footnotesize,labelfont=bf,labelsep=endash,margin=0pt]{caption} % have figure/table caption in smaller font with bf marker
\usepackage[title,header]{appendix}

\newlength{\myleftmargin} 
\newlength{\mytopmargin} 
\newlength{\myrightmargin} 
\newlength{\mybottommargin} 
\usepackage[runin]{abstract}

\newcommand{\keywords}[1]{\vspace{2mm}\noindent\textbf{Key words:} #1} % best to use within abstract
\newcommand{\pagewidetitle}[3] % to display #1=\maketitle, #2=abstract, and #3=title/author footnotes at the correct place
{%
    \twocolumn%
        [%
            \vskip-5mm%
            \begin{@twocolumnfalse}%
                #1%
                #2%
                \vspace{5mm}%
            \end{@twocolumnfalse}%
        ]%
        #3%
}

\newlength{\figurewidth}

\newcommand{\ie}{{i.e.}}

\newcommand{\eg}{{e.g.}}
\newcommand{\etal}{{et\,al.}}
\newcommand{\etc}{{etc}}

\renewcommand{\d}{\mathrm{d}}
\newcommand{\p}{\partial}

\newcommand{\e}{\mathrm{e}}
\renewcommand{\b}[1]{{\boldsymbol{#1}}} % usually italic bold (greek letter etc.)

\newcommand{\s}{\ensuremath{\text{s}}}
\newcommand{\um}{\ensuremath{\micro\metre}}%  micro meter, from siunitx
\usepackage{relsize}

\newcommand{\bmu}{\text{\textsmaller{BMU}}} % careful: if using \ensuremath, not typeset in boldface in titles

\newcommand{\oba}{\text{\textsmaller{OB}$_\text{a}$}}

\newcommand{\oca}{\text{\textsmaller{OC}$_\text{a}$}}

\newcommand{\kform}{\text{$k_\text{form}$}} % alternative to \ensuremath
\newcommand{\kres}{\text{$k_\text{res}$}}
\newcommand{\RVE}{\text{RVE}}
\newcommand{\rve}{\text{$V_\text{T}$}}
\newcommand{\bm}{\text{bm}}
\newcommand{\fbm}{\text{$f_\bm$}}
\newcommand{\sv}{\text{$s_V$}}

\begin{document}

\title{\bf Endocortical bone loss in osteoporosis:\\ The role of bone surface availability
}
\author{Pascal R Buenzli$^\text{a,1}$, C David L Thomas$^\text{b}$, John G Clement$^\text{b}$, Peter Pivonka$^\text{a}$}

\date{\small \vspace{-2mm}$^\text{a}$Faculty of Engineering, Computing \& Mathematics, The University of Western Australia, WA 6009, Australia
    \\$^\text{b}$School of Dental Science, University of Melbourne, VIC 3010, Australia
    \\\vskip 1mm \normalsize \today\vspace*{-5mm}}

\pagewidetitle{% Not strictly necessary. Remove this for standard LaTeX, or add \newcommand{\pagewidetitle}[1]{#1}
\maketitle}{

\begin{abstract}
        Age-related bone loss and postmenopausal osteoporosis are disorders of bone remodelling, in which less bone is reformed than resorbed. Yet, this dysregulation of bone remodelling does not occur equally in all bone regions. Loss of bone is more pronounced near and at the endocortex, leading to cortical wall thinning and medullary cavity expansion, a process sometimes referred to as ``trabecularisation'' or ``cancellisation''. Cortical wall thinning is of primary concern in osteoporosis due to the strong deterioration of bone mechanical properties that it is associated with. In this paper, we examine the possibility that the non-uniformity of microscopic bone surface availability could explain the non-uniformity of bone loss in osteoporosis. We use a computational model of bone remodelling in which microscopic bone surface availability influences bone turnover rate and simulate the evolution of the bone volume fraction profile across the midshaft of a long bone. We find that bone loss is accelerated near the endocortical wall where the specific surface is highest. Over time, this leads to a substantial reduction of cortical wall thickness from the endosteum. The associated expansion of the medullary cavity can be made to match experimentally observed cross-sectional data from the Melbourne Femur Collection. Finally, we calculate the redistribution of the mechanical stresses in this evolving bone structure and show that mechanical load becomes critically transferred to the periosteal cortical bone.

\end{abstract}
}% end of \pagewidetitle. Remove this for standard LaTeX
\protect\footnotetext[1]{Corresponding\mbox{ }author.\mbox{ }Email\mbox{ }address: \\\texttt{pascal.buenzli@uwa.edu.au}}

\section{Introduction}
It is well known that cortical porosity of bone increases in osteoporosis leading to a reduction in bone stiffness and strength, ultimately increasing the risk of fracture~\cite{compston,feik-etal-1997,iwamoto-etal,seeman,bell-clement-etal,thomas-feik-clement-2005,thomas-feik-clement-2006}. The temporal evolution of this deterioration process in single individuals is still poorly understood. Most of our current knowledge is deduced from cross-sectional data collected by micro-computed tomography of bones. These data indicate that changes in cortical porosity are not uniformly distributed across the cortical thickness. In particular, in long bones, loss is more pronounced near the endocortical surface~\cite{feik-etal-1997,bell-clement-etal,thomas-feik-clement-2005,thomas-feik-clement-2006,iwamoto-etal,compston,seeman,zebaze-etal}. Age-related bone loss is therefore characterized by an expansion of the marrow (or medullary) cavity. The result of this process is often referred to as ``trabecularisation'' of cortical bone, since the aged cortical bone exhibits morphological similarities with trabecular bone~\cite{zebaze-etal}. In the following, we refer to this non-uniform bone loss of cortical bone as \emph{endocortical bone loss}. Endocortical bone loss results in a reduction in cortical wall thickness and in an increase in porosity which consequently reduces the load-bearing capacity of bone~\cite{seeman}. 

Several mechanisms have been hypothesised to drive bone loss in osteoporosis, including hormonal changes, reduced physical activity, and an evolving bone micro-structure leading to changes in bone surface availability. Hormonal changes are associated with increased remodelling activity, reduction in osteoblast activity, reduction in osteoblast number, and bone imbalance within single basic multicellular units (BMUs)~\cite{manolagas,seeman,feng-mcdonald}. To explain the non-uniformity of age-related bone loss, RB~Martin suggested that bone loss may be influenced by the microscopic availability of bone surface~\cite{martin-1972,martin-1984}. Based on experimental data, Martin demonstrated a remarkable relationship between bone porosity and bone specific surface (i.e., surface area per volume of bone tissue). Using this relationship, Martin found from a simple mathematical model that in cortical bone increased resorption activity of bone cells would lead to an increase in surface area which in turn would lead to an acceleration in resorption activity, hence, creating a morphological (or geometrical) feedback~\cite{martin-1972}. In trabecular bone this situation is reversed: increased resorption activity would lead to a decrease in surface area which in turn would lead to a reduction in resorption activity. Martin's hypothesis that bone remodelling in osteoporosis contains such a morphological feedback by the evolving bone micro-structure is difficult to study experimentally and Martin did not provide any comparison to experimental data. Indeed, the time scales involved in osteoporosis are large, and there is no easy experimental control of the strength of the morphological feedback, nor of how bone morphology changes with age.

In this paper we propose a computational modelling approach suited to the investigation of the non-uniformity of bone loss in osteoporosis attributed to a morphological regulation. Changes in cortical bone volume fraction occurring in the midshaft of human femur bone are evolved numerically assuming an initial bone state and an osteoporotic condition. Our computational model follows Martin's idea~\cite{martin-1972,martin-1984}: we hypothesise that the experimentally observed non-uniform loss of bone in osteoporosis (\eg, more pronounced near and at the endocortex) is due to the superposition of:
\begin{itemize}
    \item[(i)] Hormonal changes and/or changes in the overall loading, leading to a uniform baseline of bone loss across the cortical bone;
    \item[(ii)] An evolving bone microstructure, that locally increases or decreases bone remodelling activity (by morphological feedback), and so locally increases or decreases bone loss compared to the hormonal/mechanical baseline.
\end{itemize}
Our main hypothesis is that the above mechanisms are able to explain (at least in part) the rate of endocortical bone loss represented by the expansion of the medullary cavity, for which there are experimental data available~\cite{feik-etal-1997}. The predominant loss of bone in the endocortical region is expected to increasingly transfer mechanical loading towards the periosteum.

To test these hypotheses we extend the (purely temporal) mathematical model initially proposed by Martin~\cite{martin-1984} to include a spatial component. The resulting spatio-temporal description enables us to follow the evolution of the distribution of bone volume fraction across the midshaft of a long bone. The feedback of bone morphology on the bone cells is implemented using the phenomenological relationship between bone porosity and bone specific surface mentioned above. Furthermore, we investigate changes in the distribution of mechanical stresses across the midshaft attributable to the evolving bone microstructure. The distribution of the mechanical stresses is calculated using an extension of classical beam theory to materials of non-uniform composition. This theory and the mathematical model of bone remodelling are presented in Section~\ref{sec:methods}. The numerical results are shown in Section~\ref{sec:results} and discussed in Section~\ref{sec:discussion}.

\section{Methods}\label{sec:methods}

In this section, we first present a mathematical model of bone remodelling well-suited to investigate the role of non-uniform bone surface availability for endocortical bone loss such as that which occurs leading to osteoporosis. We then present the method by which the redistribution of the internal mechanical stresses will be calculated when bone has evolving non-uniform properties.

\subsection*{Computational model of bone remodelling including morphological regulation}
\begin{figure*}[!tbp]
    \centering
    \begin{tabular}{ll}
        \small (a)&\small (b)
        \\\raisebox{0.11\height}{\includegraphics[width=\figurewidth,clip]{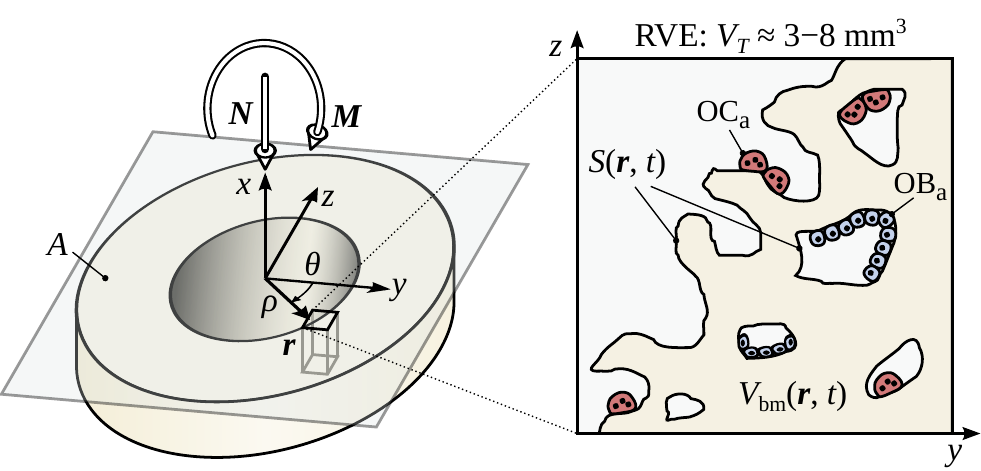}}
        &\resizebox{0.85\figurewidth}{!}{\makebox[\figurewidth]{\input{fig-sv-fbm-data}}}
    \end{tabular}
    \caption{(a) Compressive force $\b N$ and bending moment $\b M$ acting in a cross-section (of area $A$) of a long bone. The representative volume element ($\RVE$) of the tissue at position $\b r$ serves to define local spatial averages of bone properties, such as cell densities, bone matrix volume fraction $\fbm$ and specific surface $\sv$. (b) Relationship between specific surface $\sv$ and bone matrix volume fraction $\fbm$ as in Eq.~\eqref{sv-fbm} (redrawn from Ref.~\cite{martin-1984}).}\label{fig:rve}
\end{figure*}

Bone remodelling is a complex metabolic process that involves regulations at several time scales and length scales. To follow the evolution of bone properties that are relevant to the macroscopic degradation of bone in osteoporosis, we consider a computational model focused on a millimetre-size tissue scale (see Figure~\ref{fig:rve}a). This observation scale is large enough for local resorption and formation processes to be of macroscopic significance for the mechanical and microstructural properties of the tissue~\cite{hellmich-ulm-dormieux,grimal-etal}, and yet small enough for the cellular origin of resorption and formation to be considered with biochemical and morphological regulations~\cite{pivonka-etal1,pivonka-buenzli-dunstan-bookchapter}.  At the tissue scale, the various cell populations involved in bone remodelling can be represented by continuous variables, \ie\ local cell densities $n(\b r, t)$, by means of local averages over a so-called `representative volume element' ($\RVE$) of the tissue $\rve\approx 3\text{--}8\ \mm^3$:
\begin{align}
    n(\b r, t) \equiv \frac{N(\b r, t)}{\rve},
\end{align}
where $N(\b r, t)$ is the number of cells in the volume $\rve$ centred at position $\b r$, at time $t$ (see Figure~\ref{fig:rve}a).

Stiffness properties of a millimetre-size portion of bone tissue are determined to a great extent by the bone volume fraction (or equivalently the bone porosity) and the elastic properties of the mineralised bone matrix~\cite{hellmich-ulm-dormieux,grimal-etal}.\footnote{Generally speaking, pore shape and orientation are other important factors determining stiffness properties of porous materials. In cortical bone, however, pores are fairly uniformly aligned cylindrical Haversian canals and their spatial distribution within cross-sections, for example, is unimportant for tissue stiffness at the millimetre scale~\cite{grimal-etal}.} In osteoporosis, both bone volume fraction and the overall degree of mineralisation of bone matrix are decreased. Osteoporosis is associated with increased remodelling activity with net bone loss per remodelling event that becomes progressively stronger with time~\cite{seeman}. This rapid remodelling both exacerbates bone loss and replaces more densely mineralised matrix with younger, less mineralised matrix~\cite{boivin-meunier}. In this paper, we focus on age-related changes in the morphological rather than mineral properties of the bone tissue. Such morphological changes have been studied extensively in cadaver bone specimens from the Melbourne Femur Collection~\cite{feik-etal-1997,bell-clement-etal,thomas-feik-clement-2005,thomas-feik-clement-2006}. Changes in the degree of mineralisation of bone matrix will be considered in future studies. The bone volume fraction $\fbm(\b r, t)$ of a portion of bone tissue represents the relative amount of bone matrix (volume $V_\bm$) in the $\RVE$ of volume $V_T$:\footnote{The bone volume fraction $\fbm$ is also equal to $1-\Phi$, where $\Phi$ is the bone porosity. In Ref.~\cite{pivonka-etal1}, $\fbm$ was denoted by~$\text{\textsmaller{BV}}$.}
\begin{align}
    \fbm(\b r, t) \equiv \frac{V_\bm(\b r, t)}{\rve}.
\end{align}

Local bone matrix is continually renewed by remodelling. In homeostasis, resorption and formation are balanced and remodelling leads to bone turnover without loss or gain.\footnote{In cortical bone, remodelling generates both type I osteons, associated with a new Haversian canal, and type II osteons, associated with a pre-existing Haversian canal~\cite{robling-stout,parfitt-in-recker}. The creation of a new Haversian canal in type I osteons therefore always implies a net bone loss~\cite{martin-burr-sharkey}. However, the exact proportion of type I over type II osteons in normal remodelling is controversial. Whilst the density of vascular channels increases progressively with age~\cite{parfitt-in-recker}, age-related bone loss is associated with increased pore area rather than increased pore density~\cite{thomas-feik-clement-2006}. The importance of cortical bone loss with age due to normal remodelling is thus difficult to estimate. In the present model, we do not take this effect into account and we will assume the same baseline of bone imbalance per remodelling event in osteoporosis in cortical and trabecular bone for simplicity.} In osteoporosis, less bone is reformed than is resorbed, so bone turnover is associated with a net bone loss. The evolution of the local bone volume fraction $\fbm$ is governed by (i) the local densities of active osteoblasts ($\oba$) and active osteoclasts ($\oca$) and (ii) these cells' activity rates:
\begin{align}
    \tfrac{\p}{\p t} \fbm(\b r, t) = \kform \oba(\b r, t) - \kres \oca(\b r, t), \label{fbm-evol1}
\end{align}
where $\kform$ is the volume of new bone synthesised per unit time by a single active osteoblast and $\kres$ the volume of bone resorbed per unit time by a single active osteoclast. 

Equation~\eqref{fbm-evol1} expresses generally the balance between bone formation and bone resorption: $\kform \oba$ corresponds to the formation rate (bone volume fraction synthesised per unit time) and $\kres \oca$ corresponds to the resorption rate (bone volume fraction resorbed per unit time). The difficulty in Eq.~\eqref{fbm-evol1} lies in determining the evolution of the populations of active osteoclasts and active osteoblasts, and in particular, in determining how their densities depend themselves on the presence of bone matrix. Osteoclasts and osteoblasts usually require a pre-existing bone substrate to conduct their resorbing and synthesising activities.\footnote{An exception is the growth of so-called `membrane bone', which forms \emph{de novo} without cartilaginous substrate or scaffold in the flat plates of the skull.} The densities of osteoclasts and osteoblasts thus depend on the local amount of bone surface (area $S$) available to them in the $\RVE$, \ie, on the specific surface $\sv(\b r,t)$:\footnote{We do not include surfaces of the canalicular system in this definition of the specific surface $\sv$. Canalicular surfaces are not available to osteoclasts and osteoblasts for remodelling.}
\begin{align}
    \sv(\b r, t) \equiv \frac{S(\b r, t)}{\rve}.
\end{align}

Comprehensive cell population models of osteoblasts and osteoclasts including biochemical coupling have been proposed in the literature~\cite{komarova-etal,lemaire-etal,pivonka-etal1,buenzli-etal-moving-bmu,buenzli-etal-anabolic,pivonka-buenzli-specific-surface}. In Ref.~\cite{pivonka-buenzli-specific-surface}, the potential influence of the specific surface $\sv$ on various stages of osteoblast and osteoclast developments is investigated in the bone cell population model of Refs~\cite{pivonka-etal1,buenzli-etal-anabolic}. Here, however, we will consider a simpler model of bone cells, originally formulated by Martin~\cite{martin-1984}. The advantage of this simpler model is that the dependence upon $\sv$ of the densities of active osteoclasts and active osteoblasts appears explicitly. Whilst full biochemical coupling between cells is not explicit in this model, the non-uniformity and evolution of the bone surface availability are accounted for, which are important to capture non-uniform effects of remodelling during age-related bone loss according to our hypotheses.

Bone tissues with a large specific surface exhibit more remodelling, and so contain more active osteoblasts and active osteoblasts, than tissues with a small specific surface (Figure~\ref{fig:rve}a)~\cite{martin-1984}. Denoting by $\lambda_\oba$ and $\lambda_\oca$ the fractions of the bone surface at which there is osteoblastic and osteoclastic activity due to remodelling, and by $\sigma_\oba$ and $\sigma_\oca$ the surface densities of active osteoblasts and active osteoclasts at remodelling sites, the densities of active osteoblasts and active osteoclasts are given by $\oba = \lambda_\oba \sigma_\oba \sv$ and $\oca = \lambda_\oca \sigma_\oca \sv$. Substituting these expressions in Eq.~\eqref{fbm-evol1}, one thus has~\cite{martin-1984}:
\begin{align}
    \tfrac{\p}{\p t}\fbm(\b r, t) = \big( \kform \lambda_\oba \sigma_\oba - \kres \lambda_\oca \sigma_\oca \big)\ \sv(\b r, t).\label{fbm-evol2}
\end{align}
Whilst Eq.~\eqref{fbm-evol2} holds generally, in the following, it is assumed that $\kform$, $\kres$, $\lambda_\oba$, $\lambda_\oca$, $\sigma_\oba$, $\sigma_\oca$ can be taken uniform and constant in time. This is clearly a simplification, as the microscopic availability of bone surface may influence the recruitment and development of bone cells in a nontrivial way. These recruitment and development processes determine the fractions $\lambda_\oba$ and $\lambda_\oca$, which would thus normally be expected to depend additionally on $\sv$~\cite{pivonka-buenzli-specific-surface}. Here, a simple extensivity of cell densities in $\sv$ is assumed when taking $\oba$ and $\oca$ just proportional to~$\sv$. 

In osteoporosis, a net bone loss is associated with each remodelling event. To retrieve physiological overall rates of bone loss, Martin assumed a small imbalance between formation and resorption, setting~\cite{martin-1984}:
\begin{align}
    \kform\lambda_\oba\sigma_\oba-\kres\lambda_\oca\sigma_\oca = - 2\ \um\text{/year}. \label{eq-op-imbalance}
\end{align}
This constant baseline of bone loss means that a 2 $\um$-thick layer of bone matrix is resorbed each year on all bone surfaces. This baseline of bone loss is weighted by the specific surface to correspond to volumetric losses in the local $\RVE$ of the tissue $\rve$.

Bone tissues exhibit a wide range of microstructures each characterising a particular volume fraction and specific surface. Cortical bone typically has  volume fractions $\fbm\approx 0.85\text{--}0.95$ with pores mainly constituted of Haversian canals. Trabecular bone typically has volume fractions $\fbm\approx 0.15\text{--}0.55$ with `pores' corresponding to the marrow space around the trabecular plates and struts. Based on measurements performed in a large variety of bone tissues, Martin~\cite{martin-1984} has proposed that bone volume fraction $\fbm$ and specific surface $\sv$ follow an `intrinsic' or `universal' relationship, approximated by the following polynomial:
\begin{align}
    \sv(\fbm) =  a_1 \fbm + a_2 \fbm^2 + a_3 \fbm^3 + a_4 \fbm^4 +a_5 \fbm^5\label{sv-fbm},
\end{align}
with
\begin{align}
    &a_1 = 14.1/\mm,\ \ a_2 = - 10.5/\mm,\ \ a_3 = - 17.8/\mm,\notag
    \\&a_4 = 43.0/\mm,\ \  a_5 = - 28.8/\mm.
\end{align}
The relation~\eqref{sv-fbm} is plotted in Figure~\ref{fig:rve}b together with experimental data assembled in Ref.~\cite{martin-1984} from various types of human bones (femur, iliac crest, vertebra, rib) both in health and disease.\footnote{This figure is redrawn from Ref.~\cite{martin-1984} with the bone volume fraction $\fbm$ in lieu of the porosity $1-\fbm$. The coefficients $a_1,..., a_5$ of the phenomenological polynomial have been slightly adjusted from~\cite{martin-1984} such that $\sv(0) = \sv(1) = 0$.} Importantly, the specific surface exhibits a maximum $\sv^\star\approx 4.2/\mm$ at a bone volume fraction $\fbm^\ast \approx 0.63$, intermediate between cortical and trabecular bone. 

Using the phenomenological relation~\eqref{sv-fbm} and Eq.~\eqref{eq-op-imbalance} in Eq.~\eqref{fbm-evol2}, one is able to calculate the evolution of bone volume fraction $\fbm(\b r, t)$ in every region $\b r$ of the bone tissue, given an initial distribution of bone volume fraction $\fbm(\b r, t_0)$. This evolution enables us to determine how the medullary cavity expands in time for comparison with experimental data~\cite{feik-etal-1997}. It is well-known that the distinction between cortical and trabecular bone is not straightforward in osteoporosis~\cite{zebaze-etal}. Similarly, measuring the extent of the medullary cavity relies on a visual or morphological threshold to define the endocortical surface. For the purpose of this paper, we assume that the medullary cavity is defined by the region of bone tissue in which bone volume fraction is lower than a threshold $\fbm^\ast$. We will take this threshold as the bone volume fraction at which the specific surface is maximum according to Eq.~\eqref{sv-fbm}, i.e. $\fbm^\ast \approx 0.63$. The effect of choosing different threshold values for the determination of medullary cavity radius will also be investigated (see Figure~\ref{fig:results1}d).

The simplicity of the model considered here (as opposed to more detailed models of bone cell developments such as that of Ref.~\cite{pivonka-buenzli-specific-surface}) has two practical consequences: (i) the specific surface $\sv$ enters as an explicit dependence in the right hand side of Eq.~\eqref{fbm-evol2}. This allows the formulation of a semi-analytical solution for the evolution of bone volume fraction $\fbm$ (see Appendix~\ref{appx:semi-analytic-sol}); (ii) the governing equation for the bone volume fraction is self-contained: the state of the system depends only on the current distribution profile of the bone volume fraction. This enables us to easily regress the system backwards in time and to deduce the past distribution of bone volume fraction that has led to an experimentally observed medullary cavity expansion (see Appendix~\ref{appx:time-reversal}).

\subsection*{Internal mechanical stress distribution}
Mechanical loading is carried non-uniformly by bone tissues. In particular, the cortex carries the majority of the loads applied to a bone~\cite{zebaze-etal}. Here, we consider a long bone subject to a compressive resultant force $\b N(t) = - N(t)\, \hat{\b x}$ along $x$ and a resultant bending moment $\b M(t) = M_y(t)\, \hat{\b y} + M_z(t)\, \hat{\b z}$ in the cross-sectional $\b y=(y,z)$ plane (see Figure~\ref{fig:rve}a; $\hat{\b x}$, $\hat{\b y}$, and $\hat{\b z}$ denote units vectors along the $x$, $y$, and $z$ axes). To estimate how internal mechanical stresses are redistributing across the midshaft section of a long bone during age-related endocortical bone loss, we use a simple theory of elasticity extended to beams of non-uniform composition (see, \eg, Refs~\cite{timoshenko-goodier,bauchau-craig,hjelmstad}). This theory is based on two main assumptions:
\begin{enumerate}
    \item Hooke's law is valid locally at the tissue scale, \ie, the stress tensor $\sigma(\b r, t)$ and strain tensor $\varepsilon(\b r, t)$ are related by a local stiffness tensor $E(\b r, t)$ of the tissue:
    \begin{align}
        \sigma(\b r, t) = E(\b r, t) \varepsilon(\b r, t)\label{hooke}
    \end{align}
    \item Near the midshaft of a long bone, small deformations generated by compression and bending keep the initial cross-sections planar and normal to the neutral axis. This geometrical constraint on the deformations is called the \emph{Euler--Bernoulli kinematic hypothesis}.
\end{enumerate}
From a structural point of view, long bones near the midshaft can be regarded as weakly deformed long beams. In absence of torsional loads or twisting along the beam axis, as assumed here, the deformation state of such beams is well approximated by the Euler--Bernoulli hypothesis~\cite{timoshenko-goodier,hjelmstad,bauchau-craig}. The Euler--Bernoulli hypothesis implies that the strain tensor $\varepsilon(\b r, t)$ reduces to the single nonzero scalar component $\varepsilon_{xx}(\b r, t)$~\cite{hjelmstad,bauchau-craig}. For an orthotropic material such as bone~\cite{hellmich-ulm-dormieux}, this implies by Eq.~\eqref{hooke} that there are no shear stresses. The only nonzero components of the stress tensor for such materials are the normal stresses $\sigma_{xx}$, $\sigma_{yy}$ and $\sigma_{zz}$. The normal stresses $\sigma_{yy}$ and $\sigma_{zz}$ are induced by the `Poisson effect' (\ie, thickening of a material under compression) and are usually small except in particular materials such as rubber~\cite{hjelmstad,bauchau-craig},\cite[\S 17]{landau-lifshitz}. The normal stresses $\sigma_{yy}$ and $\sigma_{zz}$ do not participate directly to the transfer of the resultant compressive force $\b N(t)$ and bending moment $\b M(t)$ across the tissue. Indeed, the resultant force $\b N(t)$ and moment $\b M(t)$ are given as integrals of the stress distribution $\sigma_{xx}(\b r, t)$ in the midshaft cross-section~\cite[Secs~5.3, 6.2]{bauchau-craig} (see below). For this reason, in the following we will focus on the distribution of normal stresses $\sigma_{xx}(\b r, t)$ only. From Hooke's law~\eqref{hooke}, one has $\sigma_{xx}(\b r, t) = E_{1111}(\b r, t)\varepsilon_{xx}(\b r, t)$, and so only the single component $E_{1111}$ of the stiffness tensor needs to be considered, which corresponds to the compressive stiffness.\footnote{The calculation of the other normal stresses $\sigma_{yy}$ and $\sigma_{zz}$ depends on other components of the stiffness tensor. Using bone's orthotropic stiffness property~\cite{hellmich-ulm-dormieux} and the fact that only $\varepsilon_{xx}\neq 0$: $\sigma_{yy}=E_{1122} \varepsilon_{xx}$ and $\sigma_{yy} = E_{1133}\varepsilon_{xx}$.} We will omit any indices for notational simplicity, and write $E_{1111}(\b r, t) \equiv E(\b r,t)$.

We consider a cross-section at a position $x$ along the bone axis, near the midshaft, and use the coordinates $\b y \equiv (y,z)$ to denote a position in the cross-sectional plane (see Figure~\ref{fig:rve}a). By definition, the resultant force $\b N(t)$ and resultant  moment $\b M(t)$ are given as integrals of the stress distribution $\sigma(\b y, t)\equiv \sigma_{xx}(\b r, t)$ in the midshaft cross-section (we omit both the cross-section position `$x$' and the component indices `$xx$' from the notation). With Hooke's law~\eqref{hooke}, one thus has~\cite[Secs~5.3, 6.2]{bauchau-craig}:
\begin{align}
    &N(t) = \int_A \!\! E(\b y, t)\, \varepsilon(\b y, t)\,\d y\d z,\label{resultant-force}
    \\&M_y(t) = - \int_A \!\! z\,E(\b y, t)\, \varepsilon(\b y, t)\, \d y\d z,
    \\&M_z(t) = \int_A \!\!  y\,E(\b y, t)\, \varepsilon(\b y, t)\,\d y \d z,\label{resultant-moment}
\end{align}
where the integrals are carried over the bone cross-sectional area $A$ (Figure~\ref{fig:rve}a). The geometric constraint imposed by the Euler--Bernoulli condition implies that the strain distribution $\varepsilon(\b y, t)$ is a linear function of the transverse spatial coordinates $\b y=(y, z)$ (see \eg\ Ref.~\cite[Sec.~5.2]{bauchau-craig}):
\begin{align}
    \varepsilon(\b y, t) \equiv \varepsilon_1(t) - \kappa_3(t)\,y + \kappa_2(t)\,z\label{euler-bernoulli},
\end{align}
where $\varepsilon_1$ is the sectional axial strain, and $\kappa_2$ and $\kappa_3$ are the sectional beam curvature about the $y$ and $z$ axes, respectively. At each time $t$, the Euler--Bernoulli condition~\eqref{euler-bernoulli} together with the constraints~\eqref{resultant-force}--\eqref{resultant-moment} form a system of three linear equations determining the three unknowns $\varepsilon_1(t)$, $\kappa_2(t)$ and $\kappa_3(t)$. The coefficients of this system of equations involve so-called `sectional stiffness coefficients', \ie, zero order, first order and second order moments of the non-uniform stiffness across the section~\cite[Sec.~6.2]{bauchau-craig}:
\begin{align}
    &S(t) \equiv \int_A\!\!\! E(\b y, t)\,\d y\d z,\quad S_j(t) \equiv \int_A\!\!\! x_j\,E(\b y,t)\,\d y\d z,\notag
    \\&I_{jk} \equiv \int_A\!\!\! x_j x_k\,E(\b y, t)\,\d y\d z, \quad i,j=2,3\label{inertial-moments}
\end{align}
(with $x_2=y$ and $x_3=z$). Once $\varepsilon_1(t)$, $\kappa_2(t)$ and $\kappa_3(t)$ are known, the mechanical stress distribution in the cross-section is given by use of Hooke's law:
\begin{align}
    \sigma(\b y, t) = E(\b y, t)\ \big[\varepsilon_1(t) - \kappa_3(t)\,y + \kappa_2(t)\, z\big]
\end{align}
The sectional stiffness coefficients~\eqref{inertial-moments} entering the system of equations to solve are specified at each time $t$ provided the stiffness coefficient $E(\b y, t)$ is known. Measurements of the compressive stiffness of bone in relation to the amount of bone matrix suggest that stiffness and bone volume fraction are related by a nonlinear phenomenological relationship~\cite[and Refs cited therein]{carter-hayes77,currey,martin-burr-sharkey,vanOers-etal2008}, which we take here as:
\begin{align}
    E(\b y, t) = E\big(\fbm(\b y, t)\big) = C\ \fbm(\b y, t)^3,
\end{align}
where $C\approx 15\text{ GPa}$ and $\fbm(\b y, t)$ is governed by the balance equation~\eqref{fbm-evol2}.

\section{Results}\label{sec:results}
Equation~\eqref{fbm-evol2} determines the evolution of the bone volume fraction $\fbm(\b r, t)$ at each position $\b r$ within bone from a given initial condition $\fbm(\b r, t_0)$.  Here for simplicity only rotation-symmetric initial distributions are considered in a cross-section at $x$, implying that $\fbm$ stays rotation-symmetric at all times: $\fbm(\b r,t) = \fbm(y,z,t) = \fbm(\rho,t)$, where $\rho=\sqrt{y^2+z^2}$ is the radial coordinate. The initial condition at time $t_0$ is assumed to correspond to the bone state at the onset of osteoporosis. The time elapsed since onset of osteoporosis is denoted by $\Delta t_\text{OP}\equiv t\!-\!t_0$. A constant compressive force $N(t) \equiv 0.4$~kN and constant bending moment around the $z$ axis $M_z(t) \equiv 2~\text{kN}\cdot\text{m},\, M_y(t) \equiv 0$, are assumed when calculating the redistribution of the mechanical stresses. Below, we first show how the bone volume fraction $\fbm(\rho,t)$ evolves by means of Eq.~\eqref{fbm-evol2} from a typically expected radial profile, and we determine the medullary cavity expansion that this evolution induces (Section~\ref{sec:result-evolve}). In a second numerical experiment, we calculate the ``inverse problem'', \ie, we use morphological data by Feik~\etal~\cite{feik-etal-1997} compiled into an experimentally observed average medullary cavity expansion in men to determine what initial bone state would have generated this medullary expansion (Section~\ref{sec:result-reverse}).

\subsection{Evolution of a typical bone volume fraction distribution}\label{sec:result-evolve}
\begin{figure*}[!htbp]
    \centering
    \begin{tabular}{ll}
        (a) & (d)
        \\\makebox[\figurewidth]{\input{fig-example-fbm-rho-at-times-0-10-20-30-40-50}}
        &\makebox[\figurewidth]{\input{fig-example-medul-radii-vs-age}}
        \\(b) & (e)
        \\\makebox[\figurewidth]{\input{fig-example-stress-distributions-vs-rho}}
&\hspace{4mm}\includegraphics[width=0.9\figurewidth]{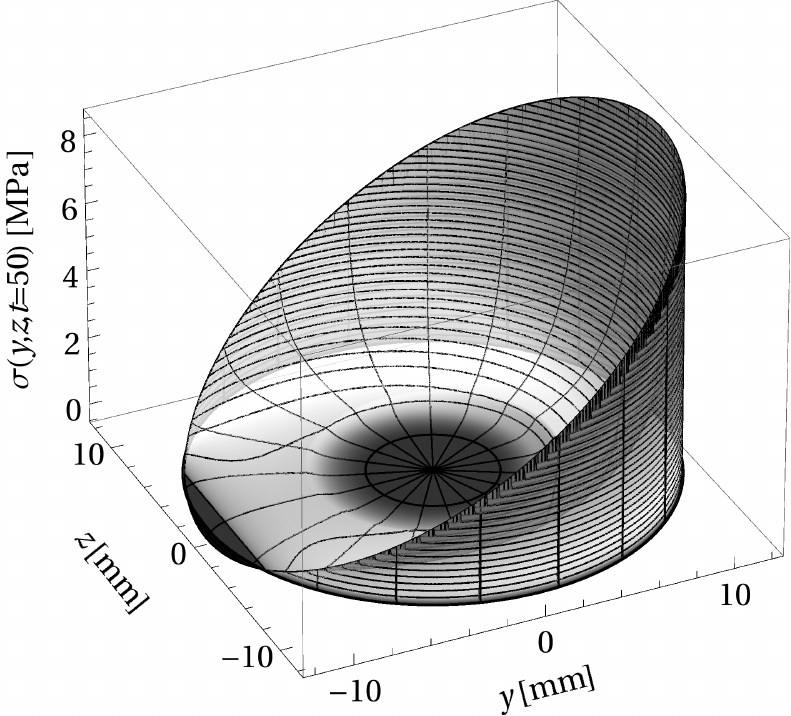}
        \\(c) & (f)
        \\\makebox[\figurewidth]{\input{fig-example-sV-rho-at-times-0-10-20-30-40-50}}
        &\hspace{13mm}\includegraphics[width=0.7\figurewidth]{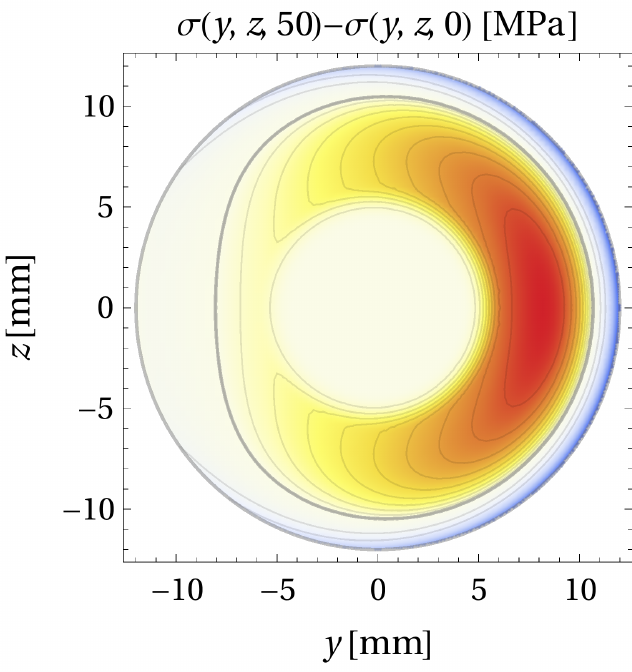}
    \end{tabular}
\caption{(a)--(c) Calculated radial profiles shown every 10 years since onset of osteoporosis for (a) bone volume fraction; (b) internal stresses at three different polar angles; (c) specific surface. The arrows indicate the evolution of the profiles with osteoporosis; (d) Calculated expansion of the medullary cavity radius. The effect of the threshold used to defined the radius is shown. Thresholds from $\fbm=0.1$ to $\fbm=0.9$ are used (thin solid lines) in addition to the threshold $\fbm=f_\bm^\ast$ at which $\sv$ is maximum (thick solid line). The filled region is indicative of the reduction in cortical wall thickness; (e) Stress distribution across the midshaft cross-section for compression and bending, exhibiting a neutral axis at $y\approx-10\ \mm$; (f) Contour plot of the difference in internal stresses between intial bone state and bone state reached after 50 years of osteoporosis. %
Red tones: decrease in stresses; Blue tones: increase in stresses; Thick grey curve: no change in stresses (zero isoline).
}
\label{fig:results1}
\end{figure*}
In Figure~\ref{fig:results1}a, the bone volume fraction radial profile $\fbm(\rho,t)$ is shown every 10 years from the onset of osteoporosis ($\Delta t_\text{OP}\!=\!0$) for 50 years. The assumed initial bone volume fraction profile at $\Delta t_\text{OP}\!=\!0$ (solid line) consists of (i) a sigmoid function of $\rho$ with value zero for $\rho\lesssim 3.3\ \mm$ (medullary cavity), sharply increasing (endocortical region) to the value $0.75$ at $\rho\approx 6.3\ \mm$ and (ii) a linear function more slowly increasing (intracortical region) to the value $0.99$ from $\rho\approx 6.3\ \mm$ to $\rho=12\ \mm$ (periosteum, bone radius). Figure~\ref{fig:results1}b and~\ref{fig:results1}c show the corresponding changes in the radial profile of internal stresses $\sigma(\rho,\theta,t)$ and specific surface $\sv(\rho,t)$ across the midshaft. In contrast to $\fbm$ and $\sv$, the internal stresses $\sigma$ are not rotation-symmetric due to the action of the bending loading condition. The evolution of $\sigma$ is shown at three angles $\theta$ of the polar coordinate system ($\theta=0$,$\tfrac{\pi}{2}$, and $\pi$), where $\theta$ is measured from the $y$ axis (see Figure~\ref{fig:rve}a). The full polar distribution of stresses is shown in Figure~\ref{fig:results1}e at $\Delta t_\text{OP}=50$~years.

The bone volume fraction in Figure~\ref{fig:results1}a is seen to decrease in all regions of the bone cross-section due to the simulated osteoporotic condition. However, as expected from Eq.~\eqref{fbm-evol2}, bone loss is more pronounced in the endocortical-to-intracortical region, where the specific surface is high. High bone volume fractions near the periosteum are comparatively more preserved. In the endocortical region, the bone volume fraction profile is both lowered and shifted towards the periosteum by the progression of osteoporosis, resulting overall in a more gradual increase of bone volume fraction from the medullary region to the periosteum than the initial profile.

The radial profile of the specific surface (Figure~\ref{fig:results1}c) gradually evolves such that its maximum originally at $\rho\approx 5.5\ \mm$ is shifted towards the periosteum. The position of this maximum corresponds to the intersection between the bone volume fraction and the constant value $\fbm^\ast\approx 0.63$ in Figure~\ref{fig:results1}a because $\fbm^\ast$ maximises the function $\sv(\fbm)$ in Eq.~\eqref{sv-fbm}. Taking $\fbm^\ast$ as the threshold value to define the transition between medullary cavity and cortex, this shift of the maximum of the specific surface towards the periosteum corresponds to an expansion of the medullary cavity radius with age. This expansion is plotted separately in Figure~\ref{fig:results1}d (thick solid line). Thin solid lines in Figure~\ref{fig:results1}d correspond to the expansion of a medullary radius defined by other threshold values of $\fbm$ as indicated by the labels. In Figure~\ref{fig:results1}d, the onset of osteoporosis was assumed at 40 years of age. For the threshold value $\fbm^\ast$, the increase in medullary radius is initially slow, then accelerates at age~$\approx 55$. The cortical wall thickness corresponds to the distance between periosteal bone radius (here assumed fixed at 12~mm) and medullary cavity radius, and is shown by the grey-shaded area. The expansion of the medullary cavity is therefore associated to a thinnning of the cortical wall thickness.

%\clearpage
These changes in morphological parameters are associated with a gradual transfer of the internal mechanical stresses towards the periosteum for $\theta=0$ and $\theta=\pi/2$ (Figure~\ref{fig:results1}b), whilst the stress radial profile at $\theta=\pi$ changes little. Interestingly, Figure~\ref{fig:results1}b suggests that for each polar angle $\theta$, there is a well-defined radius below which stresses are decreased and above which stresses are increased. The polar dependence of this radius can be appreciated in Figure~\ref{fig:results1}f. In Figure~\ref{fig:results1}f, isolines of the difference between initial stress distribution and stress distribution after 50 years of osteoporosis are shown. The radius below which stresses are decreased and above which stresses are increased corresponds to the intersection between $\sigma(\rho,\theta,\Delta t_\text{OP}\!=\!0)$ and $\sigma(\rho,\theta,\Delta t_\text{OP}\!=\!50)$ in Figure~\ref{fig:results1}b, and to the zero isoline in Figure~\ref{fig:results1}f, shown as a thick grey line. The region where stresses are increased is shaded with blue tones and the region where stresses are decreased is shaded with yellow--red tones.

%\clearpage
\subsection{Determination of past bone volume fraction profile from an observed medullary cavity expansion}\label{sec:result-reverse}
\begin{figure*}[!tbp]
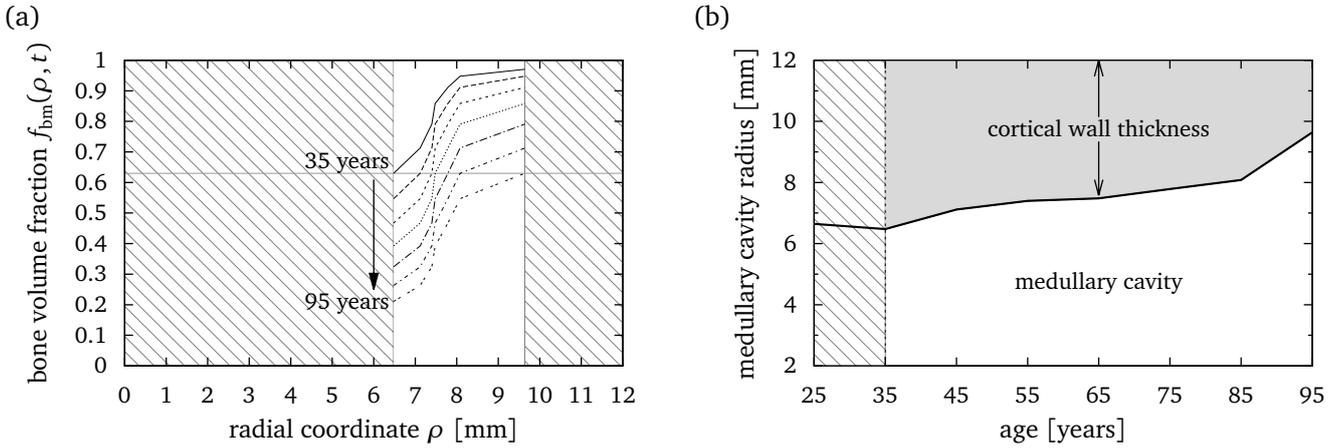

    \centering
    \begin{tabular}{ll}
        (a) & (b)
        \\\makebox[\figurewidth]{\input{fig-mfc-fbm-rho-at-times-0-10-20-30-40-50-60-men}}
        &\makebox[\figurewidth]{\input{fig-mfc-medul-radius-expansion-men}}
    \end{tabular}
    \caption{(a) Determination of bone volume fraction profiles in the endocortical region (non-hatched area) from a given expansion of medullary cavity radius (b). The data in (b) is taken from male bone samples in Ref.~\cite[Fig.~5]{feik-etal-1997}. (The parallel periosteal expansion found in these male bone samples is not considered; a constant periosteal bone radius of 12~mm is assumed here instead.) Only the monotonously increasing part of the curve (non-hatched area) can be used to determine the evolution of bone volume fraction profiles (see Appendix~\ref{appx:time-reversal}).}
    \label{fig:results2}
\end{figure*}

As explained in Appendix~\ref{appx:time-reversal}, it is possible to use our model to regress the system backwards in time. This can be used to take as input a given medullary cavity expansion and deduce from it the past distribution of bone volume fraction that leads to such an expansion when evolved by the model. Mathematically, this ``inverse problem'' can only be solved provided the medullary cavity radius is monotonously increasing. Also, the bone volume fraction profile can only be determined between the minimum and the maximum medullary cavity radii of the given expansion. In Figure~\ref{fig:results2}b, we show an observation of medullary cavity expansion obtained by Feik~\etal~\cite{feik-etal-1997} from cross-sectional samples of the Melbourne Femur Collection. In the original data, samples were grouped in decade age categories (such as 31--40 years, 41--50 years \etc.) and the medullary area was measured on each sample. Here we have transformed medullary area into medullary radius by assuming a rotation-symmetric medullary cavity and we have interpreted the value in a decade age category as the value at a single age in the timeline (namely, the middle age of the category: 31--40 years $\to$ 35 years). Because the ``inverse problem'' requires a monotonously increasing radius, we took the data presented in Figure~5 of Ref.~\cite{feik-etal-1997} for males and discarded the first data point, as indicated by the hatched area in Figure~\ref{fig:results2}b.

In Figure~\ref{fig:results2}a, we show the bone volume fraction radial profile at 35~years of age (solid line) that reproduces the medullary cavity radius of Figure~\ref{fig:results2}b when evolved to 95~years (double dashed line). As in Figure~\ref{fig:results1}a, the calculated bone volume fraction profile is shown every 10 years and the medullary cavity radius is determined by the intersection of the bone volume fraction profile with the constant value $\fbm^\ast$ (dotted horizontal line). In Figure~\ref{fig:results2}b, the minimum radius is $\approx 6.4\ \mm$ and the maximum radius is $\approx 9.6\ \mm$, and so the bone volume fraction profile in Figure~\ref{fig:results2}a is only determined between these radii as indicated by the excluded hatched areas.

\section{Discussion}\label{sec:discussion}
The study of age-related endocortical bone loss in humans is a challenging problem, with a multitude of possible causes, including biochemical, biomechanical and morphological regulations of bone remodelling. Whilst micro-CT imaging technology can give detailed insights into the bone microstructure at a given moment, to follow the evolution of bone microstructure across a whole cross-section with a degree of detail and a frequency that are sufficiently informative (i) is not currently possible without harmful radiation and (ii) would require several decades of study. Furthermore, soft tissues and bone cells are not seen in micro-CT bone scans. Two different approaches have been undertaken by biologists. Animals models of osteoporosis have been proposed, for which both micro-CT scans and histomorphometric analyses of bone samples can be performed, providing both bone morphological parameters and biochemical information. However, particularly in small animals that lack secondary osteons (such as mice and rats), the pathogenesis of osteoporosis may be quite different from that in humans. In humans, loss of endocortical bone in osteoporosis is believed to progress by expansion and coalescence of Haversian pores near the endosteum~\cite{bell-clement-etal,thomas-feik-clement-2005,thomas-feik-clement-2006}. A second approach is to collect cross-sectional data from bone samples (either \emph{ex vivo} or \emph{post mortem}) and reconstruct pseudo time sequences by grouping the data by age categories~\cite{iwamoto-etal,feik-etal-1997,bell-clement-etal,thomas-feik-clement-2005,thomas-feik-clement-2006}. The advantage of this approach is the direct measurement of human morphological properties of bone. The disadvantage is that measurements can often only be done at a single point in time in an individual.

To our knowledge, there is currently no experimental work specifically studying the influence of the microscopic bone morphology on the development and/or activity of bone cells.\footnote{Some effects of the curvature of the bone surface on osteoblast activation and/or osteoblast activity are known to occur. \emph{In vivo}, the refilling rate of cortical \bmu s depends on the current radius of the closing cavity as evidence by double tetracycling labelling techniques~\cite{lee-1964,manson-waters,mertz-etal}. In an \emph{in-vitro} setup, the local curvature of the bone substrate was also shown to influence the rate of bone formation~\cite{rumpler-fratzl-etal}.} Such a morphological feedback on bone remodelling is challenging to investigate experimentally for the following reasons: (i) a morphological feedback cannot be inhibited to estimate its influence (in contrast to biochemical regulations that can be selectively inhibited by gene knock-outs in animal models); (ii) there is no easy control of the bone microstructure \emph{in vivo}; (iii) in osteoporosis, changes in bone volume fraction occur over long time scales. Computational modelling is a powerful tool that enables the testing of hypotheses for the underlying mechanisms responsible for age-related endocortical bone loss. Most importantly, computational modelling enables the extrapolation of data measured at a single point in time into a predicted evolution over several decades. This is of particular importance for the development of patient-specific assessment and treatment tools for osteoporosis.

The main result of this paper is the observation that a morphological feedback of the bone microstructure (specifically, of the bone specific surface) on the bone cells is able to explain (at least partly) the predominant loss of bone in the endocortical region. The non-uniformity of this loss can develop from a uniform baseline of bone imbalance per remodelling event as caused by systemic hormonal changes or overall changes in mechanical loading due to reduced physical activity. The loss of endocortical bone leads to an expansion of the medullary cavity that is compatible with experimentally observed rates. Furthermore, the deterioration of bone in the endocortical region leads to a transfer of the mechanical stresses towards the periosteum.

The evolution of the radial profile of bone volume fraction in the bone midshaft in Figure~\ref{fig:results1}a is governed by a partial differential equation in time only. However, spatial non-uniformities enter the equation by means of the local availability of bone surface embodied by the specific surface $\sv$. Age-related cortical bone loss results in both an overall decrease in cortical bone volume fraction, and a decrease in cortical wall thickness due to pronounced bone loss at the endocortical surface, where the specific surface is high. This is consistent with the experimental findings of Ref.~\cite{iwamoto-etal}. 

The precise form of the expansion of the medullary cavity radius shown in Figure~\ref{fig:results1}d depends strongly on the assumed initial bone volume fraction profile and on the threshold value of bone volume fraction used to define the boundary between medullary cavity and cortex. The initial slow increase in medullary radius in Figure~\ref{fig:results1}d is due to the loss of bone in the region $3.3\ \mm \lesssim \rho \lesssim 6\ \mm$ (initial endocortical region). In this region, the intial bone volume fraction increases sharply from 0 to 0.75 with the radial coordinate $\rho$, forming a ``step'' (Figure~\ref{fig:results1}a). Once the sloped ``plateau'' of bone volume fraction for $\rho \gtrsim 6\ \mm$ (initial intracortical region) has reached the threshold bone volume fraction defining the medullary cavity boundary, the expansion of medullary radius increases rapidly in Figure~\ref{fig:results1}d before gradually slowing down due to the relative preservation of bone volume fraction near the periosteum. In fact, choosing a different threshold volume fraction essentially delays or advances this general behaviour. From these results, one can deduce that for the influence of a morphological feedback, the sharper the transition between medullary cavity (zero bone volume fraction) and cortex (high bone volume fraction), the slower the medullary cavity expansion. This observation emphasises the importance of building up ``good bones'' during growth, with high intracortical volume fractions~\cite{rizzoli-etal}.

We note that an acceleration of the medullary cavity expansion is seen in males between the age groups 81--90 and 91--100 in the data reported in Ref.~\cite[Fig.~5b]{feik-etal-1997} (corresponding to the age interval 85--95 in Figure~\ref{fig:results2}b). This acceleration could be explained by the specific shape of the bone volume fraction radial profile at the onset of osteoporosis. From our model, the bone volume fraction at age 35 shown in Figure~\ref{fig:results2}a, calculated such that it leads to the experimental medullary expansion of Figure~\ref{fig:results2}b, exhibits a similar sharp ``step'' (associated with a slow medullary expansion) followed by a ``sloped plateau'' (associated with a fast medullary expansion) as in Figure~\ref{fig:results1}a. However, other physiological factors not accounted for in our model may play a role in this experimentally observed acceleration of the medullary expansion, such as a difference in hormonal imbalance with advancing age that could lead to an increased bone imbalance per remodelling event. We also note that in Figure~5 of Ref.~\cite{feik-etal-1997}, this acceleration of the medullary expansion occurs in parallel to an expansion of the subperiosteal cross-sectional area. Cortical wall thickness decreased in the male bone samples of Ref.~\cite{feik-etal-1997} between the age groups 61--70 and 81--90 (corresponding to the age interval 65--85 in Figure~\ref{fig:results2}b) due to both an increase in medullary cavity and decrease in subperiosteal cross-sectional area. However, cortical wall thickness was relatively preserved between the age group 81--90 and 91--100 (corresponding to the age interval 85--95 in Figure~\ref{fig:results2}b) as a result of periosteal apposition. In our model, changes in periosteal bone radius are not considered. Whilst bone remodelling processes are observed at the periosteum~\cite{orwoll,seeman}, periosteal apposition is thought to occur via bone modelling processes, which are not well described by our model in its current form.

Figures~\ref{fig:results1}b and~\ref{fig:results1}e clearly show that internal stresses become increasingly and disproportionately redistributed on the strongest part of the bone midshaft, \ie, towards the bone periphery. This disproportionate redistribution could put the bone integrity at increased risk. Indeed, the periosteum is a surface with low remodelling rate~\cite{orwoll,seeman}. An acceleration of microcrack generation at the periosteum can be expected, increasing the risk of a macroscopic fracture. However, this increased risk may be offset by periosteal bone apposition~\cite{feik-etal-1997,stein-etal,seeman}.

Increased stresses towards the periosteum may also account for a conservation of intracortical pore area. In the current model a mechanical feedback to form bone at the periosteum and to slow down bone imbalance at the endocortex is not explicitly taken into account.\footnote{Setting the bone imbalance to $-2~\um$/year in Eq.~\eqref{eq-op-imbalance} could be assumed to implicitly account for a mechanical feedback at the endocortex, but such an imbalance cannot account for bone formation at the periosteum.} Figure~\ref{fig:results1}f suggests that a potential mechanical feedback based on internal stresses would lead to further non-uniformities in the evolution of the distribution of bone volume fraction in the midshaft. Indeed, the region in which stresses are increased and the region in which stresses are decreased by the simulated osteoporotic condition have a rotation-asymmetric boundary as shown by the thick grey line in Figure~\ref{fig:results1}f. The mechanical feedback would be particulary weak near the mechanical neutral axis induced by the bending loading condition and seen in Figure~\ref{fig:results1}e near $y\approx -10\ \mm$. Figure~\ref{fig:results1}f also suggests that bone loss would be accelerated by a mechanical feedback in the endocortical region where stresses are decreased, concurrently to an induction of bone formation at the periosteum in this area, suggesting an outward shift of the cortex and so an evolving bone shape.

In this work, bone tissue was regarded as being solely composed of pores (vascular porosity) and mineralised bone matrix. In particular, the volume fraction of osteoid was not considered. Osteoporosis is an evolving bone disorder, that transitions from high turnover rate with low imbalance to mid-to-high turnover rate (possibly induced by secondary hyperparathyroidism) with a gradually increasing lack of formation~\cite{seeman,manolagas}. These differences in turnover rates modify the local volume fraction of osteoid. Increased turnover rate temporarily lowers the degree of mineralisation of bone, which also contributes to a reduction in the stiffness properties of bone~\cite{boivin-meunier,hellmich-ulm-dormieux}. Here, only the morphological dimension of the progression of osteoporosis was considered. We also note that osteoid makes asymmetric the availability of bone surface between osteoblasts and osteoclasts. Indeed, active osteoclasts rarely resorb bone covered by osteoid, but active osteoblasts deposit osteoid on a previously laid osteoid layer.

In conclusion, we found that the microscopic availability of bone surface within the tissue, needed for resorption and formation processes to occur, is potentially a significant factor in focal bone loss predominantly occuring near the endocortical wall in osteoporosis. Furthermore, the gradual redistribution of the internal stresses towards the periosteum is suggestive of the possibility to induce modelling responses at the periosteum by mechanical feedback. Age-related endocortical bone loss involves a variety of biochemical and biomechanical processes that were not explicitly included in the model. Our model enables the extraction of the influence of the microscopic availability of bone surface. This influence is not easily assessed experimentally. Potential differences between experimental data on the evolution of bone volume fraction and the results of the model can therefore be attributed to these other influences, in particular a biomechanical feedback inducing other non-uniformitites in the evolution of bone volume fraction.

\subsection*{Acknowledgements}
We would like to thank S.~Scheiner for fruitful discussions. The authors wish to acknowledge those who generously gave permission for tissue to be taken from the remains of a family member. Earlier studies of this material generated much of the fundamental data that has been drawn upon for this manuscript. We also wish to thank the staff of the mortuary and the donor tissue bank at the Victorian Institute of Forensic Medicine, Melbourne, Australia for their direct assistance with the collection of research material. PP gratefully acknowledges financial support from the Australian Research Council (Project number DP0988427).

\begin{appendices}% using package appendix for more flexibility.
\section{Semi-analytic implicit solution}
\label{appx:semi-analytic-sol}
The governing equation of the bone volume fraction has the form:
\begin{align}
    \tfrac{\p}{\p t} \fbm(\b r, t) = -\eta\ s_V(\fbm(\b r, t)).\label{fbm-evol-semi-analytical}
\end{align}
with $\eta = 2\um$/year. This equation can be integrated in time at each location $\b r$:
\begin{align}
    \int_{t_0}^t \!\!\!\d t \frac{(\p/\p t) \fbm}{s_V(\fbm)} = \int_{\fbm(\b r,t_0)}^{\fbm(\b r, t)} \frac{\d f}{s_V(f)} = - \eta\, (t-t_0).
\end{align}
Denoting by $g(f)$ an anti-derivative of $1/\sv(f)$, one can thus write the solution in the following implicit form:
\begin{align}
    g\big(\fbm(\b r, t)\big) - g\big(\fbm(\b r, t_0\big) = - \eta\ (t-t_0).\label{g-fbm-implicit}
\end{align}
When $\s_V(f)$ is a polynomial, one may use the representation:
\begin{align}
    \frac{1}{\sv(f)} = \sum_{i=1}^n \frac{A_i}{f-f_i},
\end{align}
where $A_i$ are (real) constants and $f_i$ the (possibly complex) roots of $\sv(f)$, with $f_1=0$ and $f_2=1$. Depending on the degree of the polynomial fit used for $\sv(f)$, these roots can be determined either numerically or analytically. For the particular fifth order polynomial proposed in Eq.~\eqref{sv-fbm} one has (with $A_i$ in millimetre):
\begin{alignat}{3}
    &f_1=0,&\quad  &f_2=1,&\quad  &f_3\approx -0.611,\notag
    \\&f_4\approx -0.55-0.7\,\text{i},&\quad &f_5\approx 0.55+0.7\,\text{i}, &\quad& \notag
    \\&A_1 \approx -2.04,&\quad &A_2\approx 0.89,&\quad &A_3\approx 0.55, \notag
    \\&A_4\approx 0.30-0.63\,\text{i},&\quad &A_5\approx 0.30+0.63\,\text{i} &\quad&
\end{alignat}
For polynomial $\sv(f)$, the function $g(f)$ has thus the general form:
\begin{align}
    g(f) = \sum_{i=1}^n A_i \ln(f\!-\!f_i) = \ln\Big(\prod_{i=1}^n (f-f_i)^{A_i} \Big),\label{g-f}
\end{align}
and the bone volume fraction $\fbm(\b r, t)$ can be determined by solving the implicit algebraic equation (using Eq.~\eqref{g-f} in Eq.~\eqref{g-fbm-implicit}):
\begin{align}
    \prod_{i=1}^n \big(\fbm(\b r, t) - f_i\big)^{A_i} = \prod_{i=1}^n \big(\fbm(\b r, t_0) - f_i\big)^{A_i}\ \e^{-\eta\ (t-t_0)}\label{fbm-implicit}
\end{align}
This equation can also provide an implicit equation for the evolution of the medullary cavity. Assuming a cylindrical geometry, the medullary cavity radius $\rho^\ast(t)$ at time $t$ is defined by $\fbm(\rho^\ast(t), t) = f^\ast$, where $f^\ast$ is the value of bone volume fraction chosen to define the endocortical wall, for example, $f^\ast =\fbm^\ast\approx 0.63$ at which the specific surface is maximum. Evaluating Eq.~\eqref{fbm-implicit} at $\rho=\rho^\ast(t)$, one thus has:
\begin{align}
    \prod_{i=1}^n \big(\fbm(\rho^\ast(t), 0) - f_i\big)^{A_i} = \prod_{i=1}^n \big(f^\ast - f_i\big)^{A_i}\ \e^{\eta (t-t_0)}\label{medullary-cavity-radius}
\end{align}

\section{Determination of the endocortical volume fraction profile at an earlier age}
\label{appx:time-reversal}
Reversing time in the governing equation for bone volume fraction~\eqref{fbm-evol-semi-analytical} provides the evolution equation
\begin{align}
    \tfrac{\p}{\p \bar t} \fbm(\b r, \bar t) = +\eta\ \sv\big(\fbm(\b r, \bar t)\big),
\end{align}
\ie, bone volume fraction increases with reversed time $\bar t = -t$. Because $\fbm$ is the only state variable in the model presented here, one may determine the history of bone volume fraction from its current profile only. When other variables participate in determining the state of the system (such as bone cell densitites in more comprehensive models of bone cell populations), this reversal is more complex and also likely to be numerically unstable.

Below, we show that it is in fact sufficient to know the expansion of the medullary cavity in time to deduce the bone volume fraction profile at an earlier age (in particular, at the onset of osteoporosis) in the endocortical region. (This property is used in Figure~\ref{fig:results2} with the evolution of the medullary cavity area estimated in Ref.~\cite[Fig.~5]{feik-etal-1997}.) For simplicity, a rotation-symmetric cylindrical bone shaft is assumed. From Eq.~\eqref{medullary-cavity-radius}, it is clear that if the medullary cavity radius $\rho^\ast(t)$ is a monotonic increasing function of $t\geq 0$ with values in the range $[\rho_0, \rho_1]$, one can solve Eq.~\eqref{medullary-cavity-radius} for $\fbm(\rho, 0)$ for all $\rho\in[\rho_0, \rho_1]$. Indeed, it is sufficient to find the (unique) time $\tilde t$ at which $\rho^\ast(\tilde t) = \rho$ and to invert the implicit algebraic equation for $\fbm(\rho,0)$ obtained by setting $t=\tilde t$ in Eq.~\eqref{medullary-cavity-radius}.

Another approach is to evolve the system backwards in time during the time period $\tilde t$ defined above, from the initial volume fraction $\fbm^\ast$ defining the endocortical wall. Indeed, this backwards evolution directly provides the bone volume fraction $\fbm(\rho, 0)$. This approach is easier to implement numerically as it does not require the knowledge of the roots $f_i$ of the phenomenological function $\sv(f)$.

\end{appendices}

% If using a .bib file to process the bibliography:
%% \clearpage %flush table before references
%% %% \bibliographystyle{plainnat} % F. Lastname. Title. Journal (italic), volume: pages, year.
%% %% \bibliographystyle{elsarticle-harv} % Lastname, F., year. Title. Journal volume, pages.
%% \bibliographystyle{apalike} % Lastname, F. (year). Title. Journal (italic), volume:pages.
%% %% \bibliographystyle{authordate1} % Lastname, F. year. Title. Journal (italic), volume (bold), pages.
%% %% \bibliographystyle{authordate2} % same as authordate 1 but journal names first letters not capitalised
%% %% \bibliographystyle{authordate3} % same as authordate 1, but using small capitals for authors
%% %% \bibliographystyle{authordate4} % same as authordate 3, but journal names first letters not capitalised
%% %% \bibliographystyle{abbrvnat} % (same as plainnat in authordate mode) F. Lastname. Title. Journal (italic), volume: pages, year.
%% %% \bibliographystyle{unsrtnat} % (same as plainnat in authordate mode, but not alphabetically sorted)
%% \bibliography{CA}

% If typesetting the bibliography manually:

\end{document}